%% file: main-R3.tex
\documentclass{article}

\input{header}
\title{A Graphical Framework for Testing Hierarchically Structured Hypothesis Families}
\input{files/authors}

\date{\today}

\begin{document}
\maketitle

\newpage

\input{files/abstract_v2}
\input{files/introduction_v2}

\input{files/preliminary}

\input{files/method}

\input{files/case_studies}

\input{files/real_example-R3}

\input{files/simulation}
\input{files/conclusion}
\input{files/appendix}

\bibliographystyle{plainnat}
\bibliography{ref_mcp}   

\end{document}

%% file: header.tex
\usepackage[round]{natbib}
\usepackage{amsmath,amssymb,amsthm}
\usepackage{booktabs}
\usepackage{color}
\usepackage{subfig}        % ok to keep if you already use it
\usepackage{graphicx}      % use graphicx instead of epsfig

\usepackage{algorithm}     % float environment for algorithms
\usepackage{algpseudocode} % algorithmicx pseudocode

\usepackage[colorlinks=true,citecolor=blue,urlcolor=blue,linkcolor=blue]{hyperref}
\usepackage[nameinlink]{cleveref}

\theoremstyle{plain}
\newtheorem{theorem}{Theorem}[section]

\theoremstyle{definition}
\newtheorem{definition}{Definition}[section]
\newtheorem{condition}{Condition}[section]
\newtheorem{example}{Example}[section]
\newtheorem{case}{Case}[section]

\theoremstyle{remark}
\newtheorem{remark}{Remark}[section]

\crefname{condition}{Condition}{Conditions}
\Crefname{condition}{Condition}{Conditions}
\crefname{definition}{Definition}{Definitions}
\Crefname{definition}{Definition}{Definitions}
\crefname{theorem}{Theorem}{Theorems}
\Crefname{theorem}{Theorem}{Theorems}
\crefname{lemma}{Lemma}{Lemmas}
\Crefname{lemma}{Lemma}{Lemmas}
\crefname{remark}{Remark}{Remarks}
\Crefname{remark}{Remark}{Remarks}
\crefname{example}{Example}{Examples}
\Crefname{example}{Example}{Examples}
\crefname{case}{Case}{Cases}
\Crefname{case}{Case}{Cases}

\newcommand{\pr}[1]{\Pr\!\left\{#1\right\}}
\newcommand\BibTeX{{\rmfamily B\kern-.05em \textsc{i\kern-.025em b}\kern-.08em
T\kern-.1667em\lower.7ex\hbox{E}\kern-.125emX}}

\usepackage[
  letterpaper,
  top=1.3in, bottom=1.3in,
  left=1.3in, right=1.3in
]{geometry}

\renewcommand{\baselinestretch}{1.5}
\usepackage{xcolor}

 \usepackage{multirow} 

%% file: files/authors.tex
\author{Zhiying Qiu\\
Biostatistics at Insmed\\
Bridgewater, NJ 08807, U.S.A.
\and
Li Yu \\
Biostatistics at Syndax Pharmaceuticals \\
New York, NY 10017, U.S.A.
\and
Wenge Guo\\
Department of Mathematical Sciences\\
New Jersey Institute of Technology\\
Newark, NJ 07102, U.S.A. \\
Email: wenge.guo@njit.edu
}

%% file: files/abstract_v2.tex
\begin{abstract}
In clinical trials, hypotheses are frequently organized into hierarchically ordered families, requiring specialized testing strategies that account for these structured relationships. Existing gatekeeping methods—including serial, parallel, and tree-structured approaches—provide important solutions but are often either too rigid or insufficiently intuitive to accommodate increasingly complex logical dependencies among hypothesis families. To address these limitations, we propose a novel family-based graphical approach that unifies the derivation and visualization of diverse gatekeeping strategies. In this framework, procedures are represented as directed, weighted graphs, where nodes correspond to hypothesis families. Two simple updating rules govern the allocation of significance levels within families and the propagation of significance levels between them. We establish that the proposed method strongly controls the familywise error rate (FWER) at a pre-specified level. Simulation studies under representative configurations indicate that the proposed procedure achieves performance comparable to hypothesis-level graphical approaches and competitive with the superchain procedure, while providing a simpler and more interpretable family-level representation. Case studies and a real clinical trial application further illustrate its flexibility and practical advantages, making it a powerful tool for managing hierarchically structured multiple testing in clinical research.
\end{abstract}


%% file: files/introduction_v2.tex
\section{Introduction}\label{s:intro} 

In clinical trial research, complex multiple testing problems often arise from hierarchically ordered objectives \citep{alosh2014advanced, tamhane2018advances}. These typically involve hypotheses grouped into families of endpoints—primary, secondary, and sometimes tertiary—that must be tested in a sequential manner \citep{dmitrienko2009multiple}. Such hierarchical structures require specialized procedures that ensure strong control of the familywise error rate (FWER) while respecting clinical priorities.

To address these challenges, \citet{maurer1995multiple} and \citet{bauer1998testing} introduced the gatekeeping strategy, in which hypotheses in later families may only be tested if earlier families satisfy predefined gatekeeping conditions. Gatekeeping procedures are broadly classified into two types. In \emph{serial gatekeeping} \citep{westfall2001optimally}, a family is tested only if all hypotheses in the preceding families are rejected. In \emph{parallel gatekeeping} \citep{dmitrienko2003gatekeeping}, a subsequent family may be tested if at least one hypothesis in the current family is rejected.

More advanced approaches were developed to address complex logical dependencies among hypothesis families. 
The tree-structured gatekeeping strategy of \citet{dmitrienko2007tree} and its extension—the mixture procedure \citep{dmitrienko2011mixtures,dmitrienko2013general}—enable flexible testing of hierarchically ordered families under intricate logical constraints. 
Tree-structured gatekeeping operates within a fixed hierarchical framework with pre-specified logical relationships, and the testing strategy strictly follows this structure. In contrast, mixture procedures are formulated under the closure principle by constructing combined intersection $p$-values through valid mixing functions applied to family-specific intersection $p$-values. This formulation permits more general parametric multiple testing procedures and greater flexibility in handling logical constraints.
However, because mixture procedures rely on the closure principle, they require evaluation of a potentially large collection of intersection hypotheses. The number of such intersections grows exponentially with the number of hypotheses (or families), which can lead to substantial computational burden in large-scale settings.

Subsequent developments include \citet{xi2014general,dmitrienko2016mixture,kordzakhia2018enhanced,wang2022managing}. To improve practical implementation, simpler stepwise gatekeeping procedures were proposed \citep{dmitrienko2006stepwise,guilbaud2007bonferroni,dmitrienko2008general}. In particular, the general multistage procedure of \citet{dmitrienko2008general} unified earlier methods into a coherent and implementable framework that is easier to communicate in clinical applications. Nevertheless, stepwise procedures typically offer less flexibility than mixture-based methods when handling highly complex logical structures.

As logical restrictions in trial objectives become increasingly intricate, effective visualization and communication tools are essential. Graphical approaches, such as those proposed by \citet{bretz2009graphical} and \citet{burman2009recycling}, provide not only intuitive representations of sequential testing strategies but also a formal mechanism for encoding multiplicity control. Within these frameworks, transition coefficients play a pivotal role: they determine how unused portions of the significance level are reallocated across hypotheses, thereby governing the propagation of significance levels and explicitly capturing the logical dependencies among testing objectives.

Building on this idea, the superchain procedure of \citet{kordzakhia2013superchain} represents families as vertices and propagates significance levels through pre-specified transition coefficients. Although elegant and flexible, this approach tests all families simultaneously, which may reduce its suitability for strictly hierarchical trial designs. To better accommodate such settings, \citet{maurer2014note} proposed a graphical method tailored to serial gatekeeping, where the graph is updated only after all hypotheses within a family are rejected, thus aligning more closely with hierarchical testing structures.

In this paper, we introduce a new \emph{family-based graphical approach} that combines the transparency of graphical methods with the flexibility of stepwise gatekeeping. The method generalizes the multistage gatekeeping framework, enabling a broader range of logical restrictions to be accommodated. For instance, it naturally addresses scenarios in which equally important families (such as co-primary endpoints) must be tested within the same layer. By operating directly at the family level, the proposed algorithms extend existing graphical approaches and enhance the transparency and interpretability of hierarchical testing strategies. 

Family-level graphical representations have appeared in clinical trial
protocols primarily for communication purposes; for example, the CANTOS
trial protocol \citep{ridker2017antiinflammatory} employed diagrams to
illustrate the hierarchical testing strategy. However, such diagrams are
typically conceptual and do not uniquely determine a valid multiple
testing procedure, as they do not encode the within-family testing
method or the rule governing the redistribution of unused significance
levels. Consequently, the same diagram may correspond to different
implementations (e.g., serial versus parallel gatekeeping).

The proposed family-based graphical framework shares this general feature
in that the graphical structure alone does not uniquely determine a valid
testing procedure. Additional specifications are required, including the
choice of within-family testing methods and the associated error-rate
function bounds that govern the redistribution of significance levels.
Our framework makes these components explicit and integrates them with
the graphical representation. As a result, once the graph and the
associated local procedures and error-rate functions are specified, they
together define a well-defined testing algorithm and its level-updating
mechanism. This formulation reduces ambiguity in interpretation and
facilitates a rigorous analysis of strong FWER control.

Conceptually, the proposed framework may be viewed as a structured subclass of the more general superchain procedures of \cite{kordzakhia2013superchain}, in that both rely on graph-based propagation of significance levels. However, the two approaches differ substantially in scope and design objectives. Superchain procedures permit iterative re-testing and flexible recycling schemes across objectives, whereas our method is formulated as a sequential multi-layer, family-based procedure with a single forward pass, fixed transition coefficients, and no iterative updating. The primary goal of our framework is to provide a regulator-friendly construction that emphasizes transparency, interpretability, and ease of communication at the family level.

From a practical standpoint, this development is particularly relevant in modern clinical trials, where study designs frequently involve multiple co-primary endpoints, composite outcomes, or key secondary objectives that are critical for regulatory approval. Existing methods are often either too rigid to accommodate such complex logical structures or too computationally intensive for routine implementation. 
Beyond visualization, our family-based graphical approach formalizes the logical flow of hypothesis testing, explicitly encodes the updating and redistribution of significance levels, and enables procedures tailored to specific study objectives and regulatory constraints. This framework assists statisticians in designing and implementing coherent multiple testing strategies, supports clinicians in interpreting testing hierarchies and decision pathways, and facilitates regulatory reviewers in evaluating the logical structure and error-control properties of confirmatory analyses. By combining theoretical rigor with operational transparency, the proposed approach helps bridge the gap between methodological sophistication and practical clinical application, thereby enhancing the interpretability and acceptance of advanced multiple testing strategies.

The remainder of the paper is organized as follows. Section~\ref{HEURISTIC} motivates the proposed approach through two heuristic examples. Section~\ref{PRELIMINARY} introduces the notation and assumptions. Section~\ref{MAIN_THEORY} presents the general sequential testing algorithm and establishes FWER control. Section~\ref{CASES} illustrates the advantages of the approach through three case studies from \citet{bretz2009graphical}. Section~\ref{EXAMPLE} provides a real data application. Section~\ref{sec:simulation} presents a simulation study evaluating the finite-sample performance of the proposed method. Section~\ref{FUTURE_WORK} concludes with a discussion of future research directions. Technical proofs are collected in the Appendix.

%% file: files/preliminary.tex
\subsection{Heuristics}\label{HEURISTIC}

\citet{bretz2009graphical} introduced a general graphical framework as a
visualization tool for Bonferroni-based gatekeeping procedures.
In particular, they described graphical implementations of truncated
Holm-type gatekeeping procedures (see \citet{bretz2009graphical},
Section~3), in which the redistribution of unused significance levels is
governed by pre-specified truncation parameters. Such graphical
representations provide an intuitive way to visualize how significance
levels are allocated and propagated across hypotheses.

For completeness, we briefly recall the truncated Holm procedure
\citep{dmitrienko2008general}, which serves as a building block in many
graphical gatekeeping strategies.

\begin{definition}[Truncated Holm procedure]
Let $H_1,\ldots,H_m$ be $m$ hypotheses with corresponding $p$-values
$p_1,\ldots,p_m$, and let $p_{(1)} \le \cdots \le p_{(m)}$ denote the
ordered $p$-values, where $H_{(i)}$ corresponds to $p_{(i)}$. For a
truncation fraction $0 \le \gamma \le 1$ and overall significance level
$\alpha$, the truncated Holm procedure compares $p_{(i)}$ with
\[
w_i\alpha
=
\left(
\frac{\gamma}{m-i+1}+\frac{1-\gamma}{m}
\right)\alpha,
\qquad i=1,\ldots,m .
\]
Starting from $i=1$, reject $H_{(i)}$ if $p_{(i)} \le w_i\alpha$ and
continue sequentially until the first non-rejection occurs. When
$\gamma=0$, the procedure reduces to the Bonferroni procedure; when
$\gamma=1$, it coincides with the classical Holm step-down procedure.
\end{definition}

\begin{figure}
\begin{center}
\includegraphics[scale=0.7]{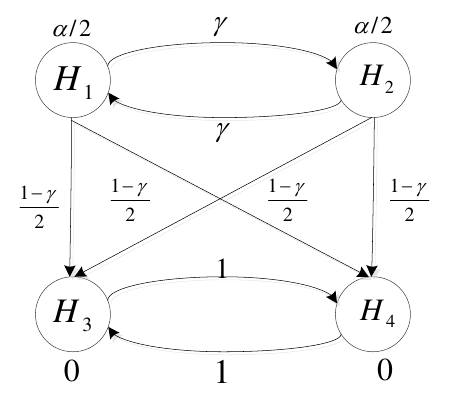}
\end{center}
\caption{Schematic graphical representation of a gatekeeping procedure
based on the truncated Holm method (after \citet{bretz2009graphical}).}
\label{TRUN_H}
\end{figure}

Figure~\ref{TRUN_H} provides a schematic illustration of a parallel
gatekeeping strategy based on a truncated Holm procedure
\citep{dmitrienko2008general}. In this example, four hypotheses are
grouped into two families, and each hypothesis is represented by a
vertex in the graph. Directed edges indicate possible transfers of
significance levels following rejections. The figure is intended to
illustrate the graphical construction rather than to uniquely represent
all possible truncated Holm parameterizations.

Compared with conventional multiple testing procedures applied to a
single family of hypotheses, hypothesis-based graphical approaches offer
a clear and flexible representation of complex testing strategies.
However, in many clinical trials multiple hierarchically ordered
families of hypotheses must be tested. As the number of families grows,
the corresponding hypothesis-level graphs may become increasingly
complex and difficult to interpret.

\begin{figure}
\begin{center}
\includegraphics[scale = 0.5]{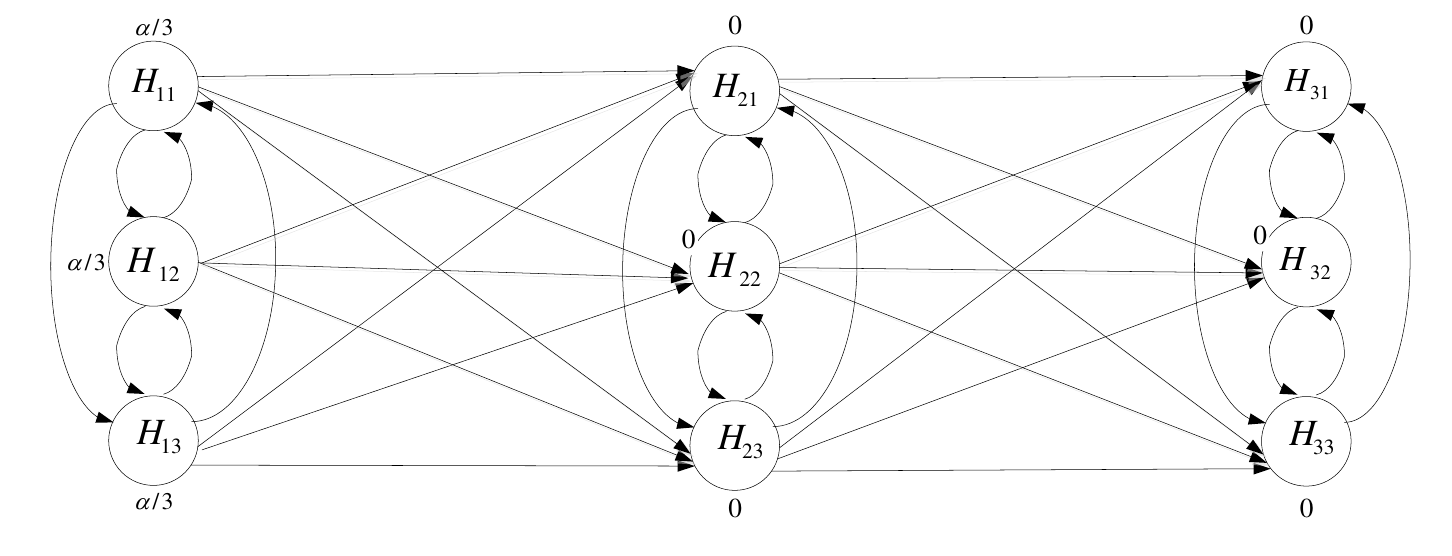}
\end{center}
\caption{Schematic hypothesis-based graphical representation of a
gatekeeping strategy using a truncated Holm procedure with truncation
parameter $\gamma$ (after \citet{bretz2009graphical}).}
\label{TH3}
\end{figure}

To illustrate this issue, consider an example with nine hypotheses
grouped into three families, each containing three hypotheses, denoted
by $F_i=\{H_{i1},H_{i2},H_{i3}\}$ for $i=1,2,3$. Suppose that $F_1$ and
$F_2$ are tested sequentially using a truncated Holm procedure with
truncation parameter $\gamma$ as described in
\citet{bretz2009graphical}, while $F_3$ is tested using the standard
Holm procedure. Furthermore, a subsequent family is tested only if at
least one hypothesis in the current family is rejected.
Figure~\ref{TH3} shows a schematic hypothesis-based graphical
representation of this strategy. To simplify the presentation, the edge
weights are omitted. The figure is intended to illustrate the structural
complexity of the hypothesis-level representation rather than to
correspond to a unique specification of truncation parameters.

In many practical applications, hierarchical logical relationships among
families of hypotheses play a central role in determining the testing
strategy. In such settings, it is often more natural and informative to
represent the structure at the \emph{family level} rather than at the
level of individual hypotheses. Motivated by the superchain framework of
\citet{kordzakhia2013superchain}, we consider a
\textit{family-based graphical representation}, in which each vertex
represents a family of hypotheses. Directed edges, together with
associated weights, describe how significance levels may be transferred
between families following rejections.

\begin{figure}
\centering
$\begin{array}{lr}
\subfloat[Strategy 1]{
\includegraphics[width=0.15\textwidth]{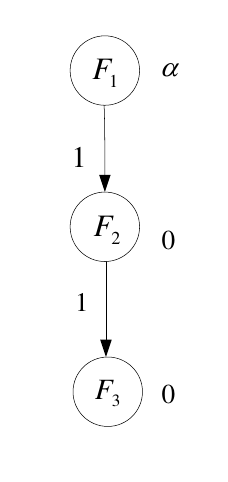}}
&
\subfloat[Strategy 2]{
\includegraphics[width=0.25\textwidth]{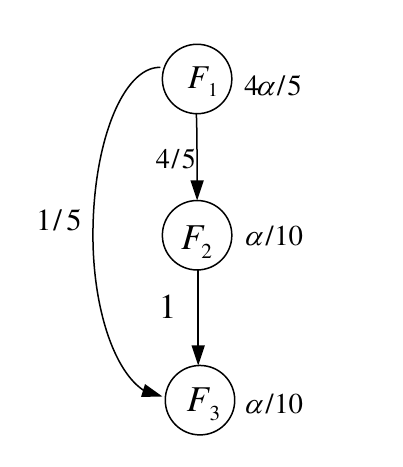}}
\end{array}$
\centering
\renewcommand{\baselinestretch}{1}
\caption{Family-based graphical visualization of gatekeeping strategies:
(a) strategy 1 and (b) strategy 2.}
\label{FM_TH1}
\end{figure}

For example, the family-based representation corresponding to the
strategy in Figure~\ref{TH3} is shown in Figure~\ref{FM_TH1}(a), where
each family $F_i$ ($i=1,2,3$) is represented as a vertex. Testing begins
with $F_1$ at significance level $\alpha$. The next family, $F_2$ (or
$F_3$), can be tested only if at least one hypothesis in $F_1$ (or $F_2$)
is rejected. Significance levels are transferred between families
through transition coefficients assigned to edges: when a rejection
occurs within a family, a portion of its significance level may be
redistributed to subsequent families according to these coefficients.
Detailed updating rules are provided in Section~3.

It is important to note that the graphical structure itself does not
uniquely determine the gatekeeping strategy. In particular, the same
graph may correspond to different implementations (e.g., serial or
parallel gatekeeping) depending on the choice of within-family testing
procedures and the associated error-rate-function bounds. Thus, the
graph should be interpreted together with these statistical components.
The representation in Figure~\ref{FM_TH1}(a) is intended to illustrate
the hierarchical relationships and a simple forward-propagation
mechanism under a specific choice of local procedures.

To further demonstrate the flexibility of this framework, consider a
parallel gatekeeping strategy in which the initial significance levels
allocated to $F_1$, $F_2$, and $F_3$ are $4\alpha/5$, $\alpha/10$, and
$\alpha/10$, respectively. In addition, $1/5$ of the significance level
from $F_1$ may be transferred to $F_3$ if at least one hypothesis in
$F_1$ is rejected. Figure~\ref{FM_TH1}(b) shows the corresponding
family-based graphical representation. In this case, the parallel
gatekeeping structure is induced by the specification of the local
testing procedures and error-rate-function bounds, rather than by the
graph alone. Even if no rejection occurs in $F_1$, the subsequent
families $F_2$ and $F_3$ may still be tested at their allocated local
significance levels.

We note that alternative edge structures (for example, allowing
additional propagation between families) can also be specified within
the same framework and, together with different choices of local
procedures, may lead to procedures with different power characteristics.
The present examples are intended primarily to illustrate the structural
representation rather than to identify a power-optimal design.

\subsection{Basic Notation and Framework}\label{PRELIMINARY}

We introduce the notation and structural assumptions underlying the proposed procedure.  
Suppose $N \ge 2$ null hypotheses are organized into $m \ge 2$ families, which are further arranged into $n$ ordered layers.

The $i$-th layer is denoted by
\[
L_i = \{F_{i1}, \ldots, F_{i l_i}\}, \qquad i = 1,\ldots,n,
\]
where $l_i$ is the number of families in layer $L_i$ and $\sum_{i=1}^n l_i = m$.
Each family $F_{ij}$ in layer $L_i$ contains $n_{ij} \ge 1$ hypotheses,
\[
F_{ij} = \{H_{ij1}, \ldots, H_{ij n_{ij}}\}, \qquad j = 1,\ldots,l_i,
\]
so that the total number of hypotheses satisfies
\[
\sum_{i=1}^n \sum_{j=1}^{l_i} n_{ij} = N.
\]

Each hypothesis $H_{ijk}$ is associated with a $p$-value $P_{ijk}$.
The overall objective is to control the familywise error rate (FWER) at a pre-specified level $\alpha$.

For any true null hypothesis, the corresponding $p$-value is assumed to be stochastically no smaller than $\mathrm{Uniform}(0,1)$.  
Specifically, letting $T_{ij}$ denote the set of true nulls in $F_{ij}$, we assume that for all $u \in [0,1]$,
\begin{equation}
\Pr\!\left(P_{ijk} \le u \mid H_{ijk} \in T_{ij}\right) \le u,
\quad \forall\, i,j,k.
\end{equation}

\begin{remark}[Definition of families and layers]
The partition of hypotheses into families and layers is assumed to be
specified prior to testing. In confirmatory clinical trials, this
structure is typically determined by the clinical objectives and
endpoint hierarchy described in the study protocol or statistical
analysis plan (SAP), rather than by statistical convenience alone.

Common grouping principles include:
(i) grouping by endpoint type (primary, key secondary, secondary),
(ii) grouping dose–placebo comparisons within the same endpoint,
(iii) placing co-primary endpoints within the same layer when they
share equal priority, and
(iv) aligning families with pre-specified gatekeeping logic in the SAP.

More generally, the relationship between hypothesis-based and
family-based representations is not one of simple equivalence. A
family-based structure can be viewed as a higher-level abstraction of a
more detailed hypothesis-based graph. To obtain a family-based
representation from a hypothesis-level graph, additional design choices
are required, including how hypotheses are grouped into families and
how transition weights are aggregated across families. These choices
are generally not unique and must be made in a way that preserves the
clinical hierarchy and yields a valid testing procedure.

The proposed framework does not prescribe a unique grouping. Instead,
it provides a rigorous testing procedure once a clinically meaningful
family structure has been specified, together with the associated
transition rules, the local testing procedures within each family, and
the corresponding error-rate-function bounds.
\end{remark}

\paragraph{Familywise error rate.}
The familywise error rate (FWER) is defined as the probability of rejecting at least one true null hypothesis. 
In the layered setting, the \emph{overall FWER} is the probability of rejecting at least one true null across all families and layers. 
A procedure is said to \emph{strongly control} the FWER at level $\alpha$ if this probability does not exceed $\alpha$ under any configuration of true null hypotheses.

\paragraph{Significance level allocation.}
Let $\alpha_i$ denote the initial significance level assigned to layer $L_i$, and let $\alpha_{ij}$ denote the portion allocated to family $F_{ij} \subset L_i$. 
These allocations must satisfy:

\begin{condition}[significance level allocation]\label{cond:allocation}
\[
\sum_{i=1}^n \alpha_i \le \alpha,
\qquad
\sum_{j=1}^{l_i} \alpha_{ij} \le \alpha_i
\quad \text{for each } i=1,\ldots,n.
\]
\end{condition}

Testing proceeds sequentially across layers $L_1,\ldots,L_n$. 
Within each layer, families may be tested in arbitrary order using their allocated significance levels. 
Unused portions of significance levels may be transferred to later families according to transition coefficients defined below.

\paragraph{Transition coefficients.}
Let $\mathbf{G}=\{g_{ijkl}\}$ denote the transition coefficients, where $g_{ijkl}$ represents the proportion of the local significance level from family $F_{ij}$ that may be transferred to family $F_{kl}$ in a subsequent layer. 
These coefficients must satisfy:

\begin{condition}[Transition coefficients]\label{cond:transition}
For all $i,j$,
\[
0 \le g_{ijkl} \le 1, 
\qquad 
g_{ijkl} = 0 \ \text{if } k \le i,
\qquad
\sum_{k=i+1}^n \sum_{l=1}^{l_k} g_{ijkl} \le 1.
\]
\end{condition}

The collections $\{\alpha_{ij}\}$ and $\{g_{ijkl}\}$ define a directed acyclic graph (DAG):
\begin{itemize}
    \item Each vertex corresponds to a family $F_{ij}$ labeled with its significance level $\alpha_{ij}$.
    \item A directed edge from $F_{ij}$ to $F_{kl}$ is present if $g_{ijkl} > 0$.
\end{itemize}
Because vertices represent families rather than individual hypotheses, we refer to this structure as a \emph{family-based graph}.

\paragraph{Error rate function.}
The updating mechanism relies on the error rate function introduced by
\citet{dmitrienko2008general}.

\begin{definition}[\citet{dmitrienko2008general}]
For a family $F=\{H_1,\ldots,H_n\}$ with an associated multiple testing
procedure, the \emph{error rate function} is defined as
\[
e(I)
=
\sup_{H_I}
\Pr\!\left(\bigcup_{i \in I} \{\text{reject } H_i\} \,\middle|\, H_I\right),
\]
for any $I \subseteq \{1,\ldots,n\}$,
where $H_I = \bigcap_{i \in I} H_i$.
\end{definition}

When $e(\cdot)$ is not available in closed form, we work with an upper
bound $e^*(\cdot)$.

\begin{example}
For the Bonferroni procedure with local level $\alpha$, the error rate
function satisfies
\[
e(I) \le \frac{|I|}{n}\,\alpha,
\qquad I \subseteq \{1,\ldots,n\},
\]
since each hypothesis is tested at level $\alpha/n$.

For the truncated Holm procedure with truncation fraction $\gamma$,
an upper bound on the error rate function is given by
\[
e^*(I)
=
\begin{cases}
0, & I = \emptyset,\\[4pt]
\bigl[\gamma + (1-\gamma)\,|I|/n\bigr]\alpha, & I \subseteq \{1,\ldots,n\},\ I \neq \emptyset.
\end{cases}
\]
In particular, $e^*(I) < \alpha$ for all non-empty $I$ when $\gamma \in [0,1)$, implying that
the procedure is separable. These bounds are used in the level-updating
mechanism.
\end{example}

\begin{remark}
In \citet{dmitrienko2008general}, the error rate function was developed
in a parallel gatekeeping setting under a separability condition
requiring $e(I_i) < \alpha_i$ unless all hypotheses in the family are
rejected. This condition ensures that the unused portion of the local
significance level is well defined for level transfer. In the present
layered framework, we do not impose this separability condition as a
general requirement. Instead, because significance levels may be
partially allocated and redistributed across families, the updating
mechanism relies directly on the error-rate-function bound (or its upper
bound $e^*(\cdot)$) rather than on separability.
\end{remark}

\paragraph{Local procedures.}
Each family $F_{ij}$ is tested using a local procedure characterized by an error rate function (or bound) $e_{ij}^*(\cdot)$. 
Let $\alpha_{ij}^*$ denote the effective significance level assigned to $F_{ij}$ during testing, and let $A$ denote the set of accepted hypotheses. 
After testing $F_{ij}$, the quantity
\[
\alpha_{ij}^* - e_{ij}^*(A)
\]
represents the unused portion of the significance level, which is redistributed to subsequent families according to the transition coefficients.

We impose the following standard conditions on local procedures.

\begin{condition}[FWER control]\label{cond:fwer}
For all $\alpha \in (0,1)$,
\[
\mathrm{FWER}(\alpha) \le \alpha.
\]
\end{condition}

\begin{condition}[Monotonicity]\label{cond:monotonicity}
For subsets $A \subseteq B$,
\[
e_{ij}^*(A) \le e_{ij}^*(B).
\]
\end{condition}

\begin{condition}[$\alpha$-consistency]\label{cond:consistency}
For $\alpha_1 \le \alpha_2$,
\[
R(\alpha_1) \subseteq R(\alpha_2),
\]
where $R(\alpha)$ denotes the rejection set at level $\alpha$.
\end{condition}

%% file: files/method.tex
\section{Methodology}\label{MAIN_THEORY}

In this section, we introduce a novel \textit{family-based graphical approach} and establish its strong control of the overall familywise error rate (FWER). We first describe the two-layer case with four hypothesis families in Subsection~\ref{TWO_FOUR}, which serves as a motivating example. The general case with multiple layers and an arbitrary number of families per layer is presented in Subsection~\ref{GEN_MUTI_LEVEL}.

The proposed family-based graphical framework relies on the availability
of an error-rate function (or a valid upper bound) for the local testing
procedure within each family. These local error-rate-function bounds are
used to determine how significance levels can be redistributed across
families while maintaining global error control.

\subsection{Two-layer case with four hypothesis families} \label{TWO_FOUR}

Consider $m=4$ families of hypotheses arranged into two layers, $L_1$ and $L_2$, with two families in each layer.  

\begin{figure}[ht]
\centering
\includegraphics[scale=0.6]{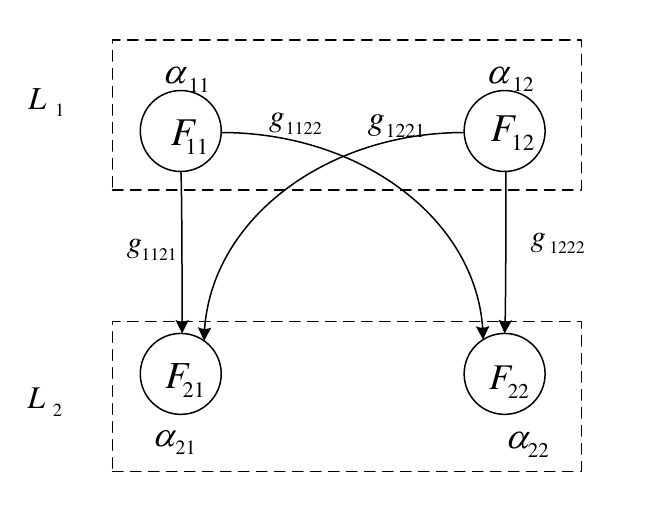}
\caption{Graphical representation of the two-layer family-based procedure with $m=4$ families of hypotheses.}
\label{TWOSTAGE}
\end{figure}

Using the notation introduced in Section~\ref{PRELIMINARY}, the procedure is described formally below.

\medskip
\noindent\textbf{Algorithm 3.1 (Two-layer family-based procedure)} \label{ALGO_TWO_LEVEL}
\medskip

\noindent\textit{Inputs.}
\begin{itemize}
  \item Layers: $L_1 = \{F_{11}, F_{12}\}$ and $L_2 = \{F_{21}, F_{22}\}$.
  \item Initial significance levels: $\alpha_{1j}$ for $F_{1j}$ ($j=1,2$) and $\alpha_{2k}$ for $F_{2k}$ ($k=1,2$).
  \item Transition coefficients: $g_{1j,2k} \in [0,1]$ with $\sum_{k=1}^2 g_{1j,2k} \le 1$ for each $j$.
  \item Local FWER-controlling procedures for each family, with error-rate bounds $e^*_{1j}(\cdot)$.
\end{itemize}

\noindent\textit{Output.}
\begin{itemize}
  \item The set of rejected hypotheses across $L_1$ and $L_2$.
\end{itemize}

\noindent\textit{Steps.}
\begin{enumerate}
  \item[\textbf{(S1)}] \textbf{First layer.} For each $j \in \{1,2\}$ (in any order):
    \begin{enumerate}
      \item Test $F_{1j}$ at level $\alpha_{1j}$ using an FWER-controlling procedure; let $A_{1j}$ denote the accepted set.
      \item Compute the unused portion
      \[
        u_{1j} = \alpha_{1j} - e^*_{1j}(A_{1j}).
      \]
      \item Update the second-layer levels: for each $k \in \{1,2\}$, set
      \[
        \alpha_{2k} \leftarrow \alpha_{2k} + u_{1j}\, g_{1j,2k}.
      \]
      \item Remove outgoing edges from $F_{1j}$ (no further transfers from $F_{1j}$).
    \end{enumerate}
  \item[\textbf{(S2)}] \textbf{Second layer.} For each $k \in \{1,2\}$ (in any order):
    \begin{enumerate}
      \item Test $F_{2k}$ at level $\alpha_{2k}$ using an FWER-controlling procedure; record the rejections.
    \end{enumerate}
\end{enumerate}

\medskip
Informally, the procedure proceeds in two stages. The families in layer~1 are tested first at their assigned levels. For each family, the portion of its significance level that is not used is redistributed to the layer~2 families according to the transition coefficients. Once both families in layer~1 have been tested, the families in layer~2 are tested at their updated levels. The overall process is illustrated in Figure~\ref{TWOSTAGE}. 

\begin{remark}
Since each family is tested only once in the proposed single-pass algorithm, outgoing edges are removed after testing. This may introduce some conservativeness compared with iterative recycling or re-testing schemes.
\end{remark}

\begin{theorem}\label{THM_TWOL_FWER}
Suppose Conditions~\ref{cond:allocation}--\ref{cond:consistency} hold.  
Then the two-layer family-based graphical procedure described in Algorithm~\ref{ALGO_TWO_LEVEL} strongly controls the overall FWER at level~$\alpha$.
\end{theorem}

A proof of Theorem~\ref{THM_TWOL_FWER} is provided in Appendix~A.1.

---

\subsection{General multi-layer family-based graphical approach} \label{GEN_MUTI_LEVEL}

The two-layer case illustrates the sequential nature of the family-based graphical procedure. We now generalize this framework to $n$ layers, each containing an arbitrary number of families.

\medskip
\noindent\textbf{Algorithm 2 (Multi-layer family-based procedure)} \label{ALGO_N_LEVEL}
\medskip

\noindent\textit{Inputs.}
\begin{itemize}
  \item Layers: $L_i = \{F_{i1}, \ldots, F_{i l_i}\}$ for $i = 1, \ldots, n$, where each layer $L_i$ contains $l_i$ families.
  \item Initial significance levels: $\alpha_{ij}$ for family $F_{ij}$, satisfying $\sum_{j=1}^{l_i} \alpha_{ij} \le \alpha_i$ for each $i$, and $\sum_{i=1}^n \alpha_i \le \alpha$.
  \item Transition coefficients: $g_{ij,kl} \in [0,1]$ for $i<k$, subject to $\sum_{k=i+1}^n \sum_{l=1}^{l_k} g_{ij,kl} \le 1$ for each $F_{ij}$.
  \item Local FWER-controlling procedures for each family, with error-rate bounds $e^*_{ij}(\cdot)$.
\end{itemize}

\noindent\textit{Output.}
\begin{itemize}
  \item The set of rejected hypotheses across all layers $L_1, \ldots, L_n$.
\end{itemize}

\noindent\textit{Steps.}
\begin{enumerate}
  \item[\textbf{(S1)}] \textbf{Layer $i$ ($1 \le i \le n-1$).} For each $j \in \{1,\ldots, l_i\}$ (in any order):
    \begin{enumerate}
      \item Test $F_{ij}$ at level $\alpha_{ij}$ using an FWER-controlling procedure; let $A_{ij}$ denote the accepted set.
      \item Compute the unused portion 
      \[
        u_{ij} = \alpha_{ij} - e^*_{ij}(A_{ij}).
      \]
      \item Update subsequent layers: for each $k>i$ and each $l \in \{1,\ldots,l_k\}$, set
      \[
        \alpha_{kl} \leftarrow \alpha_{kl} + u_{ij}\, g_{ij,kl}.
      \]
      \item Remove outgoing edges from $F_{ij}$ (no further transfers from $F_{ij}$).
    \end{enumerate}
  \item[\textbf{(S2)}] \textbf{Final layer $L_n$.} For each $j \in \{1,\ldots,l_n\}$ (in any order):
    \begin{enumerate}
      \item Test $F_{nj}$ at level $\alpha_{nj}$ using an FWER-controlling procedure; record the rejections.
    \end{enumerate}
\end{enumerate}

\medskip
Informally, the multi-layer algorithm applies the same logic as the two-layer case but extends it sequentially across $n$ layers. At each step, families in the current layer are tested at their assigned levels. For each family, the portion of its significance level not used is redistributed to families in subsequent layers according to the transition coefficients. Once all families in a layer have been tested, the procedure advances to the next layer. This process continues until the final layer is reached, at which point all remaining families are tested at their updated levels.

\begin{theorem}\label{THM_MULTI_FWER}
Suppose Conditions~\ref{cond:allocation}--\ref{cond:consistency} hold.  
Then the multi-layer family-based graphical procedure described in Algorithm~\ref{ALGO_N_LEVEL} strongly controls the overall FWER at level~$\alpha$.
\end{theorem}

A proof of Theorem~\ref{THM_MULTI_FWER} is provided in Appendix~A.2.

---

\begin{remark} \rm
Consider the special case of $n$ hierarchically ordered families, with each layer $L_i$ containing a single family $F_{i1}$. The multi-layer family-based graphical approach specializes as follows:
\begin{itemize}
    \item The initial significance level is $\alpha_{11}=\alpha$, and $\alpha_{i1}=0$ for $i>1$.  
    \item Transition coefficients are $g_{i1,\, (i+1)1}=1$ for $i=1,\ldots,n-1$, and zero otherwise.
\end{itemize}

Several implications follow:
\begin{enumerate}
\item If each local procedure controls the FWER and satisfies the
\emph{separability condition} (i.e., the error rate function is strictly
less than $\alpha$ unless all hypotheses in the family are rejected),
then the resulting level-updating mechanism in the multi-layer
family-based framework coincides with that of a parallel gatekeeping
strategy, and is equivalent to the multistage gatekeeping procedure of
\citet{dmitrienko2008general}. The separability condition is a property of
the local testing procedures rather than of a specific gatekeeping
structure. Examples of procedures satisfying this condition include the
Bonferroni, truncated Holm, and truncated fallback procedures.

\item If each family uses a local procedure with error rate function upper bound $e^*(I)=\alpha$ for any non-empty set $I$, then the multi-layer approach reduces to a serial gatekeeping strategy. Examples include the Holm and fixed-sequence procedures.  

\item If each family contains exactly one hypothesis, the procedure reduces to the conventional fixed-sequence method.  

\item When prior dependence information among $p$-values is available, additional local procedures can be used. For instance, if the null $p$-values are independent or positively dependent, one may use the Hochberg or truncated Hochberg procedures.
\end{enumerate}
\end{remark}

%% file: files/case_studies.tex
\section{Case Studies}\label{CASES}

\begin{figure}
\begin{center}
\includegraphics[scale = 0.6]{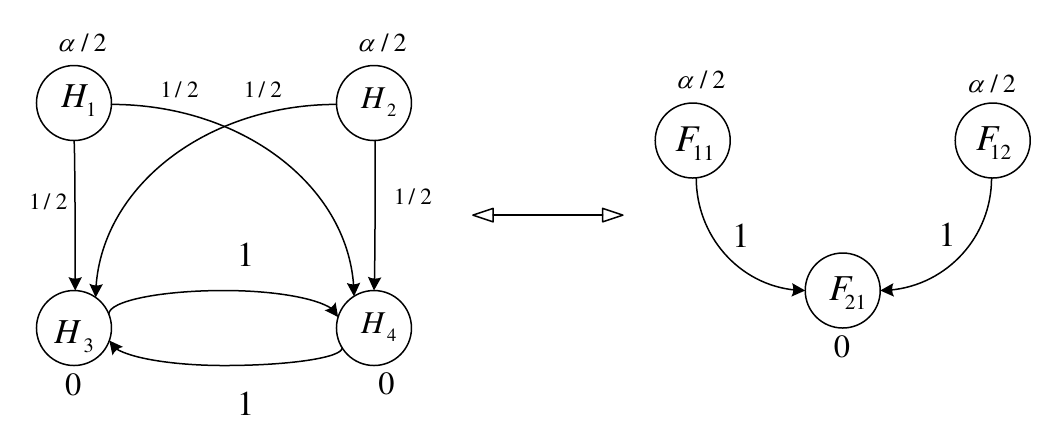}
\end{center}
\renewcommand{\baselinestretch}{1}
\caption{Graphical visualizations of Case 1: Hypothesis-based (left) and family-based (right) approaches.}
\label{EX1}
\end{figure}

\begin{figure}
\begin{center}
\includegraphics[scale = 0.6]{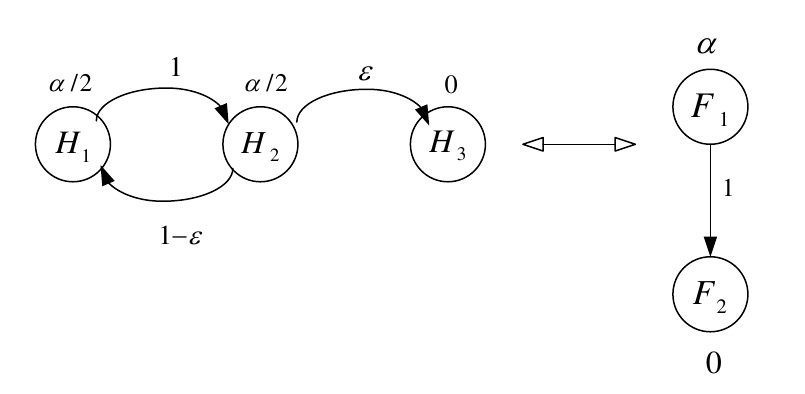}
\end{center}
\renewcommand{\baselinestretch}{1}
\caption{Graphical visualizations of Case 2: Hypothesis-based (left) and family-based (right) approaches.}
\label{EX2}
\end{figure}

\begin{figure}
\begin{center}
\includegraphics[scale = 0.6]{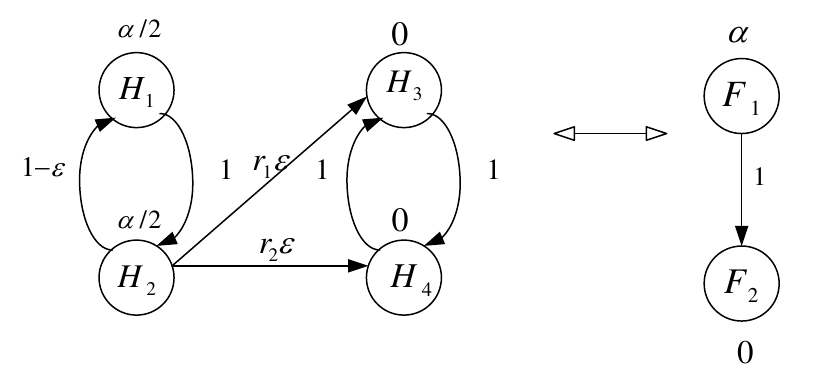}
\end{center}
\renewcommand{\baselinestretch}{1}
\caption{Graphical visualizations of Case 3: Hypothesis-based (left) and family-based (right) approaches.}
\label{EX3}
\end{figure}

In this section, we demonstrate the efficiency and simplicity of our proposed family-based graphical approach by comparing it with the conventional hypothesis-based graphical approach in the context of testing multiple families of hypotheses. Three illustrative examples from \cite{bretz2009graphical} are used for this comparison. Figures \ref{EX1}–\ref{EX3} display the original hypothesis-based graphs (left panels) alongside the corresponding family-based graphs (right panels).

\begin{case}\rm
Figure \ref{EX1} involves four null hypotheses: $H_1, H_2, H_3, H_4$. The hypothesis-based procedure is shown on the left, while the equivalent family-based procedure is shown on the right. In the family-based framework, the hypotheses are grouped into $m=3$ families: $F_{11}=\{H_1\}$, $F_{12}=\{H_2\}$, and $F_{21}=\{H_3, H_4\}$, organized across $n=2$ layers: $L_1=\{F_{11}, F_{12}\}$ and $L_2=\{F_{21}\}$. Initial significance levels are $\alpha/2$ for $F_{11}$ and $F_{12}$, and $0$ for $F_{21}$. The transition coefficient set $\mathbf{G}$ is defined as
\[
g_{1121}=g_{1221}=1, 
\quad g_{2111}=g_{2112}=g_{1112}=g_{1211}=0.
\]

The procedure begins by testing $F_{11}$ at level $\alpha/2$ using the Bonferroni method. If $H_1$ is rejected, its significance level is transferred to $F_{21}$, raising its level from $\alpha_{21}=0$ to $\alpha_{21}^*=\alpha/2$. Otherwise, no transfer occurs. Next, $F_{12}$ is tested at level $\alpha/2$; if $H_2$ is rejected, $\alpha/2$ is added to $\alpha_{21}^*$. After testing both families in $L_1$, $F_{21}$ is tested (if $\alpha_{21}^* \neq 0$) using the Holm procedure at level $\alpha_{21}^*$. This procedure is equivalent to the hypothesis-based method, but as shown in the right panel, the family-based representation more clearly conveys the hierarchical structure among families of hypotheses. \hfill $\square$
\end{case}

In some scenarios, hypotheses within one family are testable only if all hypotheses in another family are rejected. Using the hypothesis-based graphical approach to represent such strategies often leads to edges with infinitesimal weights, making the graphs unnecessarily complex and difficult for non-specialists to interpret. The family-based approach avoids this complication, as shown below.

\begin{case}\rm
Consider a gatekeeping strategy with three hypotheses: $H_1$, $H_2$, and $H_3$, where $H_3$ can only be tested if both $H_1$ and $H_2$ are rejected. The hypothesis-based graph (Figure \ref{EX2}, left) uses an edge with an infinitesimally small weight $\epsilon$. In contrast, the family-based graph (Figure \ref{EX2}, right) eliminates this edge. The strategy reduces to a two-layer, two-family procedure: $L_1=\{F_1\}$ and $L_2=\{F_2\}$, where $F_1=\{H_1,H_2\}$ and $F_2=\{H_3\}$. Initial significance levels are $\alpha$ for $F_1$ and $0$ for $F_2$. The gatekeeping procedure is:

\begin{enumerate}
\item Test $F_1$ using the Holm procedure at level $\alpha$.
\item If both $H_1$ and $H_2$ are rejected, transfer $\alpha$ to $F_2$ and test it at level $\alpha$.
\item Otherwise, stop the procedure.
\end{enumerate}
\end{case}

\begin{case}\rm
Now consider a more complex gatekeeping strategy with four hypotheses: $H_1, H_2, H_3, H_4$, where $H_3$ and $H_4$ can only be tested if both $H_1$ and $H_2$ are rejected. The hypothesis-based graph (Figure \ref{EX3}, left) again requires edges with infinitesimally small weights. Specifically, when $H_1$ and $H_2$ are both rejected, $\alpha$ is split between $H_3$ and $H_4$ according to weights $r_1$ and $r_2$, so that $H_3$ receives $r_1\alpha$ and $H_4$ receives $r_2\alpha$.

In the family-based approach (Figure \ref{EX3}, right), the structure simplifies to two layers with two families: $L_1=\{F_1\}$ and $L_2=\{F_2\}$, where $F_1=\{H_1,H_2\}$ and $F_2=\{H_3,H_4\}$. Initial significance levels are $\alpha$ for $F_1$ and $0$ for $F_2$. The procedure is:

\begin{enumerate}
\item Test $F_1$ with the Holm procedure at level $\alpha$.
\item If both $H_1$ and $H_2$ are rejected, transfer $\alpha$ to $F_2$ and test it at level $\alpha$ using a weighted Holm procedure with weights $r_1$ and $r_2$.
\item If either $H_1$ or $H_2$ is not rejected, stop the procedure.
\end{enumerate}
\end{case}

\begin{remark}\rm
Across the three examples, our family-based graphical approach consistently simplifies the representation and implementation of complex testing strategies. Unlike the hypothesis-based approach—which often requires non-intuitive infinitesimal edge weights ($\epsilon$)—the family-based approach offers a clearer and more intuitive framework, facilitating interpretation and communication, especially for non-statisticians.
\end{remark}

%% file: files/real_example-R3.tex
\section{A Clinical Trial Example} \label{EXAMPLE}

In this section, we illustrate the application of the proposed family-based graphical approach in a clinical trial setting and compare its performance with the conventional hypothesis-based graphical approach.

We revisit the Type II diabetes clinical trial example from \cite{dmitrienko2007tree}. The trial evaluates three doses of an experimental drug (Low [L], Medium [M], and High [H]) versus placebo (Plac) across one primary endpoint (P: Hemoglobin A1c) and two secondary endpoints (S1: Fasting serum glucose; S2: HDL cholesterol). Each endpoint is assessed for all three doses, yielding nine null hypotheses, which are grouped into three families:
\begin{itemize}
    \item $F_1$: dose–placebo comparisons for the primary endpoint $P$, i.e., H vs Plac ($H_{11}$), M vs Plac ($H_{12}$), and L vs Plac ($H_{13}$);
    \item $F_2$: dose–placebo comparisons for the secondary endpoint $S1$, i.e., H vs Plac ($H_{21}$), M vs Plac ($H_{22}$), and L vs Plac ($H_{23}$);
    \item $F_3$: dose–placebo comparisons for the secondary endpoint $S2$, i.e., H vs Plac ($H_{31}$), M vs Plac ($H_{32}$), and L vs Plac ($H_{33}$).
\end{itemize}

This structure highlights how the family-based graphical approach naturally organizes grouped and hierarchically ordered hypotheses.

The overall Type I error rate is controlled at $\alpha=0.05$, and the raw $p$-values for the nine hypotheses are reported in Table \ref{CH3_TABLE}. Since the primary endpoint ($P$) is considered most critical, $F_1$ is always tested first, followed by $F_2$ and $F_3$. For the secondary endpoints, we explore two types of hierarchical relationships, leading to two gatekeeping strategies: Procedure~1 and Procedure~2. Both are visualized using the family-based approach, with Procedure~1 also compared to its hypothesis-based counterpart.

\medskip

\noindent {\sc \textbf{Procedure 1.}} Suppose $S1$ and $S2$ are equally important. Then $F_2$ and $F_3$ are placed in the same layer, with hypotheses within each family tested in a pre-specified order (H vs Plac, M vs Plac, L vs Plac). The fixed-sequence procedure serves as the local test within each family. Initial significance levels are allocated as 0.04 for $F_1$, 0.005 for $F_2$, and 0.005 for $F_3$. Once $F_1$ is tested, any significance level it retains is equally split between $F_2$ and $F_3$.

\begin{figure}
\centering
$\begin{array}{cc}
\subfloat[Family-based]{
\includegraphics[width=0.3\textwidth]{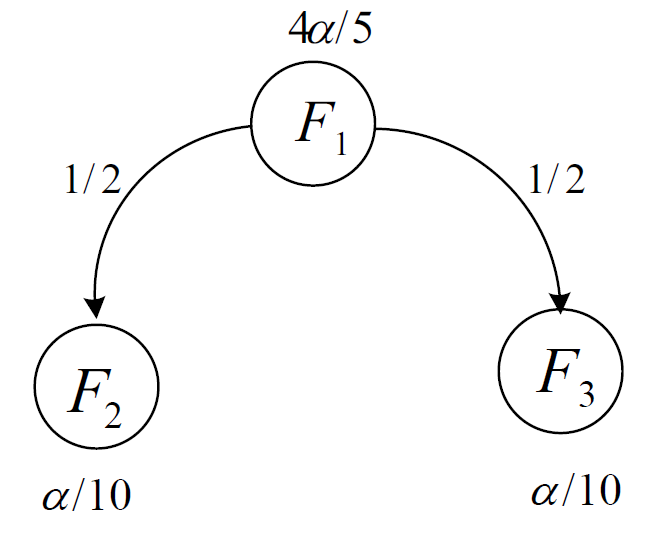}}
&
\subfloat[Hypothesis-based]{
\includegraphics[width=0.45\textwidth]{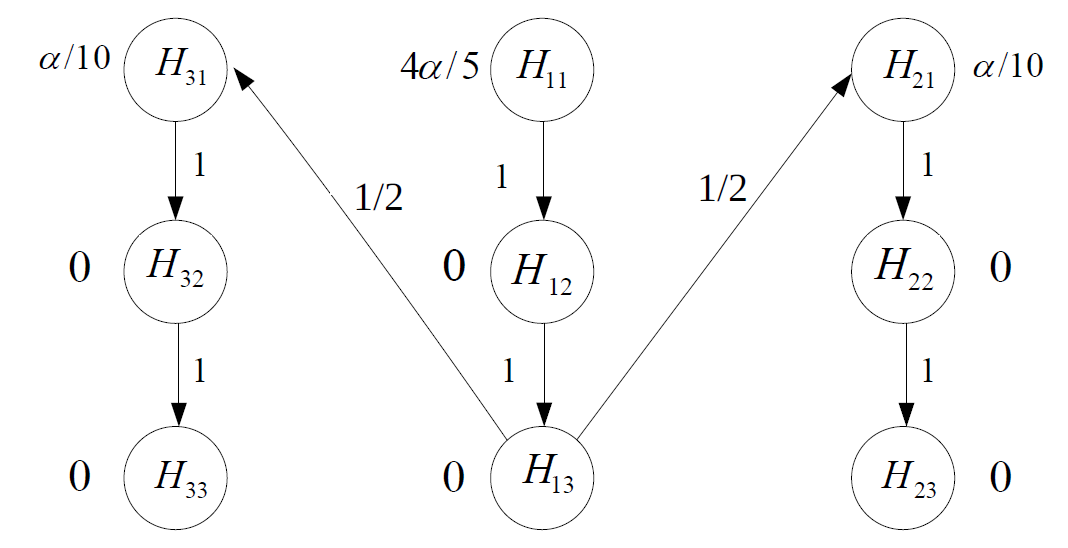}}
\end{array}$
\renewcommand{\baselinestretch}{1}
\caption{Family-based (a) and hypothesis-based (b) graphical representations of Procedure 1 for the Type II diabetes clinical trial.}
\label{TWOLEVEL}
\end{figure}

The testing proceeds as follows. At level 0.04, all three hypotheses in $F_1$ are rejected. Its significance level is then equally divided, updating the levels for $F_2$ and $F_3$ to $0.025$ each. Testing continues for $F_2$ and $F_3$ (order irrelevant), resulting in the rejection of $H_{21}, H_{31},$ and $H_{32}$. Results for Procedure~1 are summarized in Table \ref{CH3_TABLE}. Figure \ref{TWOLEVEL}(a) shows the corresponding family-based graph, while Figure \ref{TWOLEVEL}(b) displays the conventional hypothesis-based graph. The comparison illustrates that the family-based representation more transparently conveys the hierarchical structure.

\medskip

\noindent {\sc \textbf{Procedure 2.}} Suppose $S1$ is considered more important than $S2$, yielding a strict hierarchical ordering: $F_1 \to F_2 \to F_3$. The gatekeeping strategy is shown in Figure \ref{FM_TH1}(b). The truncated Hochberg procedure with truncation parameter $\gamma=0.6$ is used for $F_1$ and $F_2$, while the conventional Hochberg procedure is applied to $F_3$. Initial significance levels are 0.04 for $F_1$, 0.005 for $F_2$, and 0.005 for $F_3$.

The procedure begins with $F_1$ at level 0.04, where all hypotheses are rejected. Its significance level is then redistributed: $0.04 \times 0.8 + 0.005 = 0.037$ to $F_2$, and $0.04 \times 0.2 + 0.005 = 0.013$ to $F_3$. Next, $F_2$ is tested at level 0.037; all hypotheses are rejected, and its entire significance level is transferred to $F_3$, increasing its level to $0.05$. Finally, $F_3$ is tested at level 0.05, leading to the rejection of $H_{31}$ and $H_{32}$. Results are reported in Table \ref{CH3_TABLE}. Notably, this three-layer procedure cannot be conveniently visualized using the conventional hypothesis-based approach, underscoring the added flexibility of the family-based framework.

\begin{table}
\renewcommand{\baselinestretch}{1}
\caption{Comparison of results for two-layer (Procedure 1) and three-layer (Procedure 2) family-based procedures in the Type II diabetes clinical trial. The overall Type I error rate is $\alpha=0.05$. S = Significant; NS = Not Significant.}
\renewcommand{\baselinestretch}{2}
\begin{center}
    \begin{tabular}{|c|c|c|c|}
        \hline
       \textbf{Null hypothesis} & \textbf{Raw} $\boldsymbol{p}$\textbf{-value} & \textbf{Procedure 1} & \textbf{Procedure 2} \\ \hline\hline
        $H_{11}$        & 0.005         & S         & S          \\
        $H_{12}$        & 0.011         & S         & S          \\
        $H_{13}$        & 0.018         & S         & S          \\ \hline
        $H_{21}$        & 0.009         & S         & S          \\
        $H_{22}$        & 0.026         & NS        & S          \\
        $H_{23}$        & 0.013         & NS        & S          \\ \hline
        $H_{31}$        & 0.010         & S         & S          \\
        $H_{32}$        & 0.006         & S         & S          \\
        $H_{33}$        & 0.051         & NS        & NS         \\
        \hline
    \end{tabular}
\end{center}
\label{CH3_TABLE}
\end{table}

\begin{remark}\rm
This clinical trial example illustrates that the family-based graphical approach can provide a concise and structured representation of the testing strategy at the family level. Such a representation may be helpful for describing hierarchical relationships among families of hypotheses and for communicating the overall testing strategy.

We note, however, that this does not imply greater power or a broader class of representable procedures than hypothesis-based graphical approaches. In particular, hypothesis-based graphical approaches, possibly combined with local procedures such as Simes or Hochberg \citep{xi2019symmetric}, may yield more powerful tests in some settings. The main advantage of the family-based framework is therefore its interpretability and convenience as a family-level representation.
\end{remark}

%% file: files/simulation.tex
\section{Simulation Study}
\label{sec:simulation}

To evaluate the finite-sample performance of the proposed family-based graphical procedure, we conducted a simulation study comparing it with a hypothesis-based graphical procedure and a superchain procedure. The main objective is to assess whether the family-level representation introduces additional conservativeness relative to hypothesis-level procedures while maintaining strong control of the familywise error rate (FWER).

We considered two layers of hypothesis families,
\[
F_1=\{H_{11},H_{12}\}, \qquad F_2=\{H_{21},H_{22}\},
\]
with total significance level $\alpha=0.05$. Throughout, the initial family weights were set to
\[
w_1=1, \qquad w_2=0,
\]
so that the initial local significance levels for the two families were
\[
\alpha_{11}=w_1\alpha= \alpha,
\qquad
\alpha_{21}=w_2\alpha=0.
\]

\medskip
\noindent
\textbf{Compared procedures.}
\begin{itemize}
    \item \textbf{Proposed two-layer family-based graphical procedure.}
    The first family $F_1$ was tested using the truncated Holm procedure with truncation fraction $\gamma\in[0,1]$, while the second family $F_2$ was tested using the standard Holm procedure. The procedure operates at the family level. First, $F_1$ is tested at level $\alpha$. Let $A_1$ denote the set of hypotheses in $F_1$ that remain accepted after this step. The unused local significance level is then computed through the error-rate bound of the truncated Holm procedure,
    \[
    e_1^*(A_1)=
    \begin{cases}
    0, & |A_1|=0,\\[4pt]
    \Bigl[\gamma+(1-\gamma)\frac{|A_1|}{|F_1|}\Bigr]\alpha, & |A_1|>0,
    \end{cases}
    \]
    and the unused amount
    \[
    u_1= \alpha-e_1^*(A_1)
    \]
    is transferred to $F_2$. Since the edge weight from $F_1$ to $F_2$ was set to 1 in this study, the updated level for $F_2$ becomes
    \[
    \alpha_2^{\text{upd}}=u_1.
    \]
    The family $F_2$ is then tested using the standard Holm procedure at level $\alpha_2^{\text{upd}}$.

    \item \textbf{Hypothesis-based graphical procedure.}
This procedure is illustrated in Figure~\ref{TRUN_H}. Each hypothesis is represented as a node in a graphical multiple testing framework, with initial local significance levels
\[
\left(\frac{\alpha}{2},\, \frac{\alpha}{2},\, 0,\, 0\right)
\]
assigned to $(H_{11}, H_{12}, H_{21}, H_{22})$, respectively. Thus, the hypotheses in $F_1$ each receive level $\alpha/2$, while those in $F_2$ receive zero initial weight.

The transition weights reflect the family structure: each of $H_{11}$ and $H_{12}$ transfers a proportion $\gamma$ of its local level to the other hypothesis in $F_1$, and distributes the remaining proportion $1-\gamma$ equally between $H_{21}$ and $H_{22}$. Within $F_2$, the two hypotheses are connected by unit transition weights, corresponding to the standard Holm-type updating mechanism.

Testing proceeds sequentially according to the standard graphical multiple testing algorithm, with local significance levels updated after each rejection. In this setting, the initial configuration is equivalent to testing the hypotheses in $F_1$ using the truncated Holm procedure at level $\alpha$, while assigning zero initial significance level to the hypotheses in $F_2$. As level propagates to $F_2$, testing within $F_2$ follows the standard Holm procedure, aligning the simulation design with the other procedures.

\item \textbf{Superchain procedure.}
We consider a two-family superchain procedure with $w_1=1$ and $w_2=0$ to ensure comparability with the other procedures. Under this specification, all initial significance level is assigned to $F_1$, so that testing starts from the first family, while $F_2$ receives significance level only through subsequent updating.

In Step~1, $F_1$ is tested at level $\alpha$ using the truncated Holm procedure with truncation parameter $\gamma$, whereas $F_2$ is not tested initially. If both families still contain non-rejected hypotheses after Step~1, they are retested in subsequent steps at updated levels $\alpha_{1k}=\kappa_{1k}\alpha$ and $\alpha_{2k}=\kappa_{2k}\alpha$, where
\[
\kappa_{1k}=1,
\qquad
\kappa_{2k}=1-f_1(A_{1,k-1}\mid \gamma),
\]
and $A_{1,k-1}$ denotes the set of non-rejected hypotheses in $F_1$ after Step~$k-1$. Thus, partial rejection in $F_1$ leads to partial transfer of significance level to $F_2$, whereas complete rejection of $F_1$ allows $F_2$ to be tested at the full level $\alpha$. Throughout, $F_1$ is tested using the truncated Holm procedure, while $F_2$ is tested using the standard Holm procedure.

If only one family still contains non-rejected hypotheses, that family is retested at level $\alpha$, and the algorithm terminates. This specification retains the superchain updating structure while aligning the directional flow of significance level with that of the proposed family-based and hypothesis-based graphical procedures.
\end{itemize}

\medskip
\noindent
\textbf{Data-generating mechanism.}
For simplicity, we generated independent one-sided $z$-tests across hypotheses. Specifically, for each hypothesis we generated
\[
Z_{ij}\sim N(\mu_{ij},1),
\]
and computed the corresponding one-sided $p$-value as
\[
p_{ij}=1-\Phi(Z_{ij}).
\]
A true null hypothesis corresponds to $\mu_{ij}=0$, while a false null corresponds to $\mu_{ij}=\Delta$. In the simulations reported below, we set $\Delta=2.2$, $\gamma = 0.5$, and $\alpha = 0.05$.

\medskip
\noindent
\textbf{Simulation scenarios.}
We considered the following four configurations:
\begin{itemize}
    \item \textbf{Scenario 1:} all four hypotheses are true;
    \item \textbf{Scenario 2:} only $H_{11}$ is false;
    \item \textbf{Scenario 3:} $H_{11}$ and $H_{21}$ are false;
    \item \textbf{Scenario 4:} $H_{11}, H_{12}$, and $H_{21}$ are false, while $H_{22}$ is true.
\end{itemize}
This setup allows us to examine performance under varying signal sparsity and across-family signal configurations.

\medskip
\noindent
\textbf{Performance measures.}
For each scenario, we generated 10{,}000 Monte Carlo replicates. For each procedure, we recorded:
\begin{itemize}
    \item the familywise error rate (FWER), defined as the probability of rejecting at least one true null hypothesis;
    \item the average power, defined as the expected proportion of false null hypotheses that are correctly rejected, that is,
\[
\text{Average Power}
= \mathbb{E}\left(
\frac{\#\{\text{correctly rejected false nulls}\}}{\#\{\text{false nulls}\}}
\right).
\]
\end{itemize}
When all hypotheses are true, power is not applicable and is therefore omitted.

\begin{table}[!htbp]
\centering
\caption{Simulation results under different scenarios.}
\begin{tabular}{lcc}
\hline
\multicolumn{3}{c}{Scenario 1} \\
\hline
Procedure & FWER & Power \\
\hline
Proposed family-based graph & 0.04997 & --- \\
Hypothesis-based graph      & 0.04997 & --- \\
Superchain                  & 0.04997 & --- \\
\hline
\end{tabular}

\vspace{0.5cm}

\begin{tabular}{lcc}
\hline
\multicolumn{3}{c}{Scenario 2} \\
\hline
Procedure & FWER & Power \\
\hline
Proposed family-based graph & 0.03951 & 0.5949 \\
Hypothesis-based graph      & 0.03951 & 0.5949 \\
Superchain                  & 0.04704 & 0.5960 \\
\hline
\end{tabular}

\vspace{0.5cm}

\begin{tabular}{lcc}
\hline
\multicolumn{3}{c}{Scenario 3} \\
\hline
Procedure & FWER & Power \\
\hline
Proposed family-based graph & 0.03678 & 0.416860 \\
Hypothesis-based graph      & 0.03678 & 0.416860 \\
Superchain                  & 0.04386 & 0.418355 \\
\hline
\end{tabular}

\vspace{0.5cm}

\begin{tabular}{lcc}
\hline
\multicolumn{3}{c}{Scenario 4} \\
\hline
Procedure & FWER & Power \\
\hline
Proposed family-based graph & 0.0208 & 0.56336 \\
Hypothesis-based graph      & 0.0208 & 0.56336 \\
Superchain                  & 0.0226 & 0.58667 \\
\hline
\end{tabular}

\label{tab:simulation_results}
\end{table}

Table~\ref{tab:simulation_results} reports the results. All three procedures approximately control the FWER at the nominal level across all scenarios.

In Scenarios~2--4, the proposed family-based graphical procedure and the hypothesis-based graphical procedure yield identical performance in terms of both FWER and power, indicating that, in this setting, the family-level representation does not introduce additional conservativeness relative to the hypothesis-based graphical approach.

The superchain procedure exhibits slightly higher FWER and consistently achieves marginally higher power, particularly in Scenario~4 where most hypotheses are false. This gain in power reflects the effect of iterative updating and retesting.

The observed pattern of average power reflects a tradeoff between multiplicity and signal strength: as the number of false hypotheses increases, power may initially decrease due to the increased multiplicity burden, and then increase as the overall signal becomes stronger.

Overall, the proposed family-based graphical procedure achieves performance comparable to the hypothesis-based graphical approach, while offering a simpler and more interpretable family-level representation, with only minor differences relative to the superchain procedure.

\medskip
\noindent
We also considered the improved parallel gatekeeping procedure presented in Figure~12 of \citet{bretz2009graphical}. Additional simulations (not reported) indicate that its performance is essentially indistinguishable from that of the hypothesis-based graphical approach included here, with negligible differences in both familywise error rate control and power. Accordingly, it is omitted from the simulation tables.

\medskip
\noindent
The code used to generate all simulation results will be made publicly available upon publication of the manuscript.

%% file: files/conclusion.tex
\section{Conclusions} \label{FUTURE_WORK}

We introduce a family-based graphical framework for testing hierarchically ordered families of hypotheses. The proposed approach provides strong control of the familywise error rate (FWER) at a pre-specified level, while offering a transparent and systematic way to construct and visualize a broad class of gatekeeping strategies. As a special case, when each layer contains a single family, the framework recovers the general multistage gatekeeping procedures of \citet{dmitrienko2008general}.

The primary contribution of this work lies in providing a unifying family-based graphical framework that enables transparent construction and interpretation of hierarchical testing strategies, rather than introducing a fundamentally new testing algorithm.

Through illustrative examples, simulation studies, and a real clinical trial application, we demonstrate that the proposed framework is often simpler and more interpretable than hypothesis-level graphical approaches (e.g., \citet{bretz2009graphical}) in settings involving multiple hierarchically ordered families. The simulation results indicate that the proposed procedure achieves performance comparable to hypothesis-level graphical procedures, while remaining competitive with the superchain procedure. These findings suggest that any potential conservativeness due to family-level aggregation is scenario-dependent and remains limited in practice.

By operating at the family level, the framework aligns naturally with clinical objective structures (such as primary and key secondary endpoints), leading to clearer visualization and improved communication of complex testing strategies, particularly for non-statistical stakeholders. Overall, the proposed framework provides a practical balance between statistical rigor and interpretability, making it well suited for complex hierarchical testing problems in modern clinical trials.

Conceptually, the proposed framework is related to the superchain procedures of \citet{kordzakhia2013superchain}, which allow iterative re-testing and flexible propagation of significance levels. In contrast, our approach adopts a sequential multi-layer design with fixed transition coefficients and no iterative updating. This restriction yields a more transparent and regulator-friendly procedure, while maintaining rigorous error control and avoiding the additional complexity of more general recycling schemes.

Like any methodological framework, the proposed approach has certain limitations. In studies with many layers or a large number of families, the construction and specification of family-level graphs may become cumbersome. In addition, practical implementation in large-scale trials may require careful coordination between design-level specification and software implementation.

Another limitation is that the proposed framework relies on the availability of an error-rate function, or a valid upper bound, for the local multiple testing procedure within each family. These local error-rate-function bounds are essential for determining how significance levels can be redistributed across families while maintaining global error control. Consequently, the approach is most naturally applicable when the within-family procedures admit explicit or easily computable bounds, such as Bonferroni or truncated Holm procedures. When such bounds are not available, implementation may be more challenging, and hypothesis-based graphical approaches may offer greater flexibility in such settings.

\paragraph{Relation to hypothesis-level graphical tools.}
Hypothesis-based graphical procedures and associated software tools (e.g., gMCP and related implementations) provide flexible and widely used solutions for multiple testing in clinical trials. The proposed framework is complementary rather than competitive with these approaches. While hypothesis-level graphs offer fine-grained control, they can become difficult to construct and interpret when numerous families, layers, and within-family procedures are involved.

The family-based graphical framework operates at a higher structural level, encoding family-wise logical relationships directly and integrating explicit updating rules through error-rate-function bounds. As a result, once the family structure, transition rules, and local procedures are specified, the framework defines a well-posed and FWER-valid testing procedure, rather than serving as a purely conceptual diagram. In practice, the family-based design can be implemented using standard within-family procedures and, if desired, translated into a hypothesis-level graph for software execution. Thus, the framework provides a scalable and transparent design structure while remaining compatible with existing implementation tools.

Future work may explore extensions to adaptive designs, more complex dependence structures, and broader regulatory applications, further enhancing the applicability of the family-based graphical approach in modern confirmatory trials.

\section*{Acknowledgments}

%The research of Wenge Guo was supported in part by NSF Grant DMS-1309162. 
The authors thank the Editor, the Associate Editor, and the reviewers for their careful evaluation of the manuscript and for their insightful and constructive comments, which have significantly improved the quality and clarity of the paper.

\section*{Conflict of Interest}

The authors declare that there are no competing interests.

%% file: files/appendix.tex
\section*{Appendix}
\subsection*{A.1 \; Proof of Theorem \ref{THM_TWOL_FWER}}

We establish strong control of the FWER by partitioning the overall error event into two components:  
(i) the contribution from the first-layer families, and  
(ii) the contribution from the second-layer families, conditional on no error occurring in the first layer.

Suppose that family $F_{ij}$ is tested at level $\alpha^*_{ij}$. Then
\begin{eqnarray}\nonumber
&& \alpha^*_{1j} = \alpha_{1j}, \\[6pt]
&& \alpha^*_{2i} = \alpha_{2i} + \sum_{j=1}^{2} \bigl(\alpha^*_{1j} - e^*_{1j}(A_{1j})\bigr) g_{1j2i}, \label{UPDATE_CRITICAL_V}
\end{eqnarray}
for $i,j = 1,2$, where $e^*_{1j}$ denotes the upper bound of the error rate function for $F_{1j}$ and $A_{1j}$ is the set of accepted hypotheses in $F_{1j}$.

Let $E_{ij}(x)$ denote the event that at least one true null in $F_{ij}$ is rejected at level $x$, and let $\overline{E}_{ij}(x)$ denote its complement.  
Then the overall FWER can be decomposed as
\begin{eqnarray}\label{TWO_LEVEL_FWER}
\mathrm{FWER} &=& \pr{\bigcup_{i=1}^{2}\bigcup_{j=1}^{2}E_{ij}(\alpha^*_{ij})} \\
&=& \pr{\bigcup_{j=1}^{2}E_{1j}(\alpha^*_{1j})} \;+\; \pr{\Bigl(\bigcap_{j=1}^{2}\overline{E}_{1j}(\alpha^*_{1j})\Bigr) \cap \Bigl(\bigcup_{j=1}^{2}E_{2j}(\alpha^*_{2j})\Bigr)}. \nonumber
\end{eqnarray}

---

\textbf{Step 1: Bound for the first term.}  
For the first term in \eqref{TWO_LEVEL_FWER}, the Bonferroni inequality yields
\begin{eqnarray}
\pr{\bigcup_{j=1}^{2} E_{1j}(\alpha^*_{1j})} 
&\le& \sum_{j=1}^{2} \pr{E_{1j}(\alpha^*_{1j})} \nonumber \\
&\le& \sum_{j=1}^{2} e^*_{1j}(T_{1j}), \label{FIRST_FWER}
\end{eqnarray}
where the second inequality follows from the definition of the error rate function.  
Note that $e^*_{1j}(T_{1j})$ depends on $\alpha^*_{1j}$, since it is the error rate function of the specified multiple testing procedure at level $\alpha^*_{1j}$.

---

\textbf{Step 2: Bound for the second term.}  
Consider the second term of \eqref{TWO_LEVEL_FWER}.  
If $\bigcap_{j=1}^{2}\overline{E}_{1j}(\alpha^*_{1j})$ holds, then $T_{1j} \subseteq A_{1j}$, implying by Condition~\ref{cond:monotonicity} that
\[
e^*_{1j}(T_{1j}) \le e^*_{1j}(A_{1j}), \quad j=1,2.
\]
From \eqref{UPDATE_CRITICAL_V}, it follows that
\begin{eqnarray}\nonumber
\alpha^*_{2i} 
&=& \alpha_{2i} + \sum_{j=1}^{2}\bigl(\alpha^*_{1j} - e^*_{1j}(A_{1j})\bigr) g_{1j2i} \\
&\le& \alpha_{2i} + \sum_{j=1}^{2}\bigl(\alpha^*_{1j} - e^*_{1j}(T_{1j})\bigr) g_{1j2i}.
\end{eqnarray}
Hence, by Condition~\ref{cond:consistency},
\begin{eqnarray}
\Bigl(\bigcap_{j=1}^{2}\overline{E}_{1j}(\alpha^*_{1j})\Bigr) \cap \Bigl(\bigcup_{j=1}^{2}E_{2j}(\alpha^*_{2j})\Bigr)
\subseteq \bigcup_{i=1}^{2}E_{2i}\!\left(\alpha_{2i} + \sum_{j=1}^{2}\bigl(\alpha^*_{1j} - e^*_{1j}(T_{1j})\bigr) g_{1j2i}\right). \nonumber
\end{eqnarray}
Applying the Bonferroni inequality again gives
\begin{eqnarray}
&&\pr{\Bigl(\bigcap_{j=1}^{2}\overline{E}_{1j}(\alpha^*_{1j})\Bigr) \cap \Bigl(\bigcup_{j=1}^{2}E_{2j}(\alpha^*_{2j})\Bigr)} \nonumber \\
&\le& \sum_{i=1}^{2}\pr{E_{2i}\!\left(\alpha_{2i} + \sum_{j=1}^{2}\bigl(\alpha^*_{1j} - e^*_{1j}(T_{1j})\bigr) g_{1j2i}\right)}. \label{SECOND_FWER_0}
\end{eqnarray}

By Condition~\ref{cond:fwer}, each $F_{2i}$ is tested using an FWER-controlling local procedure.  
Thus, the probability inside the sum is bounded by its level, so the right-hand side of \eqref{SECOND_FWER_0} is at most
\begin{eqnarray}
\sum_{i=1}^{2}\left(\alpha_{2i} + \sum_{j=1}^{2}\bigl(\alpha^*_{1j} - e^*_{1j}(T_{1j})\bigr) g_{1j2i}\right).
\end{eqnarray}
Rearranging gives
\begin{eqnarray}
&=& \sum_{i=1}^{2}\alpha_{2i} + \sum_{j=1}^{2}\bigl(\alpha_{1j}- e^*_{1j}(T_{1j})\bigr)\sum_{i=1}^{2}g_{1j2i} \nonumber \\
&\le& \sum_{i=1}^{2}\alpha_{2i} + \sum_{j=1}^{2}\alpha_{1j} - \sum_{j=1}^{2} e^*_{1j}(T_{1j}) \nonumber \\
&\le& \alpha - \sum_{j=1}^{2} e^*_{1j}(T_{1j}), \label{SECOND_FWER}
\end{eqnarray}
where the first inequality follows from $\sum_{i=1}^{2}g_{1j2i} \le 1$.

---

\textbf{Step 3: Combine bounds.}  
Combining \eqref{FIRST_FWER} and \eqref{SECOND_FWER} with \eqref{TWO_LEVEL_FWER}, we obtain
\[
\mathrm{FWER} \;\le\; \sum_{j=1}^{2} e^*_{1j}(T_{1j}) + \alpha - \sum_{j=1}^{2} e^*_{1j}(T_{1j}) = \alpha.
\]

Thus, the two-layer family-based graphical procedure achieves strong control of the FWER at level~$\alpha$. \hfill $\square$

\vskip 5pt

\subsection*{A.2~~~ Proof of Theorem \ref{THM_MULTI_FWER}}

We prove the result by induction on the number of layers.

Let $\mathrm{FWER}_n(\alpha_1,\ldots,\alpha_n)$ denote the overall familywise error rate (FWER) of the multi-layer family-based procedure when the initial significance levels assigned to the layers $L_i$ are $\alpha_i$ for $i=1,\ldots,n$. 
Within each layer $L_i$, the families $F_{ij}$ receive initial significance levels $\alpha_{ij}$ for $j=1,\ldots,l_i$, subject to Condition~\ref{cond:allocation}. 
We aim to show that
\begin{equation}\label{INDUCTION}
\mathrm{FWER}_n(\alpha_1,\ldots,\alpha_n) \;\le\; \sum_{i=1}^{n}\sum_{j=1}^{l_i}\alpha_{ij} \;\le\; \alpha .
\end{equation}

\textbf{Base case ($n=2$).}
By Theorem~\ref{THM_TWOL_FWER} (proved under Conditions~\ref{cond:allocation}–\ref{cond:consistency}),
\[
\mathrm{FWER}_2(\alpha_1,\alpha_2) \;\le\; \sum_{i=1}^{2}\sum_{j=1}^{l_i}\alpha_{ij} \;\le\; \alpha .
\]

\textbf{Induction step.}
Assume \eqref{INDUCTION} holds for $n=k$ ($k\ge 2$), i.e.,
\[
\mathrm{FWER}_k(\alpha_1,\ldots,\alpha_k) \;\le\; \sum_{i=1}^{k}\sum_{j=1}^{l_i}\alpha_{ij} \;\le\; \alpha .
\]
We show it holds for $n=k+1$:
\[
\mathrm{FWER}_{k+1}(\alpha_1,\ldots,\alpha_{k+1}) \;\le\; \sum_{i=1}^{k+1}\sum_{j=1}^{l_i}\alpha_{ij} \;\le\; \alpha .
\]

\paragraph{Step 1: Decomposition.}
Define 
\[
B_1=\{\text{at least one true null is rejected in }L_1\},\quad 
B_2=\{\text{at least one true null is rejected in }L_2,\ldots,L_{k+1}\}.
\]
Then
\begin{equation}\label{eqn7}
\mathrm{FWER}_{k+1}(\alpha_1,\ldots,\alpha_{k+1})
= \pr{B_1} \;+\; \pr{\overline{B}_1 \cap B_2}.
\end{equation}
By the definition of the error-rate function and Bonferroni,
\begin{equation}\label{eqn71}
\pr{B_1} \;\le\; \sum_{j=1}^{l_1} e^*_{1j}(T_{1j}),
\end{equation}
where $T_{1j}$ is the set of true nulls in $F_{1j}$.

\paragraph{Step 2: Bounding $\pr{\overline{B}_1\cap B_2}$.}
After testing all families in $L_1$, the unused portions $\sum_{j=1}^{l_1}\left(\alpha_{1 j}-e_{1 j}^*\left(A_{1 j}\right)\right)$ are redistributed forward according to later layers. 
For any $F_{ij}$ in $L_i$ ($i\ge 2$),
\[
\alpha_{ij}^* \;=\; \alpha_{ij} + \sum_{l=1}^{l_1}\bigl(\alpha_{1l}-e^*_{1l}(A_{1l})\bigr) g_{1lij},
\]
where $A_{1l}$ is the accepted set in $F_{1l}$.
Let $\alpha_i^*=\sum_{j=1}^{l_i}\alpha_{ij}^*$.

On $\overline{B}_1$ no true null is rejected in $L_1$, so errors can only occur in $L_2,\ldots,L_{k+1}$:
\begin{equation}\label{eqn8}
\pr{\overline{B}_1 \cap B_2} \;=\; \mathrm{FWER}_k(\alpha_2^*,\ldots,\alpha_{k+1}^*).
\end{equation}
Moreover, $\overline{B}_1$ implies $T_{1j}\subseteq A_{1j}$; by Condition~\ref{cond:monotonicity},
\[
e^*_{1j}(T_{1j}) \;\le\; e^*_{1j}(A_{1j}),\qquad j=1,\ldots,l_1.
\]
Applying the induction hypothesis to \eqref{eqn8} yields
\begin{align*}
\mathrm{FWER}_k(\alpha_2^*,\ldots,\alpha_{k+1}^*)
&\le \sum_{i=2}^{k+1}\sum_{j=1}^{l_i}\alpha_{ij}^* \\
&= \sum_{i=2}^{k+1}\sum_{j=1}^{l_i}\alpha_{ij}
   + \sum_{l=1}^{l_1}\alpha_{1l}\sum_{i=2}^{k+1}\sum_{j=1}^{l_i} g_{1lij}
   - \sum_{l=1}^{l_1} e^*_{1l}(A_{1l})\sum_{i=2}^{k+1}\sum_{j=1}^{l_i} g_{1lij} \\
&\le \sum_{i=2}^{k+1}\sum_{j=1}^{l_i}\alpha_{ij}
   + \sum_{l=1}^{l_1}\alpha_{1l}
   - \sum_{l=1}^{l_1} e^*_{1l}(A_{1l})
   \qquad \text{(by Condition~\ref{cond:transition})}\\
&\le \sum_{i=1}^{k+1}\sum_{j=1}^{l_i}\alpha_{ij}
   - \sum_{j=1}^{l_1} e^*_{1j}(T_{1j})
   \qquad \text{(by Condition~\ref{cond:monotonicity}).}
\end{align*}
Here, Condition~\ref{cond:transition} ensures $\sum_{i=2}^{k+1}\sum_{j=1}^{l_i} g_{1lij}\le 1$ for each $l$. 
(Condition~\ref{cond:consistency} guarantees that increases in significance levels cannot decrease rejection sets and underpins the forward redistribution argument.)

\paragraph{Step 3: Conclusion.}
Combining \eqref{eqn7}, \eqref{eqn71}, and the bound above gives
\[
\mathrm{FWER}_{k+1}(\alpha_1,\ldots,\alpha_{k+1})
\;\le\; \sum_{i=1}^{k+1}\sum_{j=1}^{l_i}\alpha_{ij}
\;\le\; \alpha,
\]
where the final inequality uses Condition~\ref{cond:allocation}. 
This completes the induction and establishes \eqref{INDUCTION} for all $n\in\mathbb{N}$. \hfill $\square$

% \section*{Acknowledgements}
% The research of Wenge Guo was supported in part by NSF Grant DMS-1309162.